\documentclass[10pt,conference,letterpaper]{IEEEtran}

\usepackage[T1]{fontenc}
\usepackage{newtxtext,newtxmath}
\usepackage{graphicx}
\usepackage{booktabs}
\usepackage{cite}
\usepackage[caption=false,font=footnotesize]{subfig}
\usepackage{balance}
\usepackage{url}
\usepackage{xspace}
\usepackage{amsmath}
\usepackage{array}

\title{Characterizing Predictability–Throughput Trade-offs of KV-Cache SSD Offloading in LMCache for LLM Serving Systems} 

\author{
  \IEEEauthorblockN{
    \begin{tabular}{c}
       Ying He$^{\dag}$, Dingsen Shi$^{\dag}$, Yanbo Dai$^{\dag}$, Donger Chen$^{\dag}$, Harry Fu$^{\dag}$, Chris Tsu$^{\ddag}$, Wang Feng$^{\dag}$, Song Fu$^{\dag}$
    \end{tabular}
  }
  \IEEEauthorblockA{
    \text{$^{\dag}$Department of Computer Science and Engineering} \\
    \text{University of North Texas, Denton, Texas, USA}\\
    \text{\{YingHe, DingsenShi, YanboDai, DongerChen, HarryFu, WangFeng\}@my.unt.edu}, Song.Fu@unt.edu}
    \text{$^{\ddag}$Initio Corporation, San Jose, California, USA.} christsu@outlook.com\\
}

\begin{document}
\maketitle

\begin{abstract}
KV-cache offload is widely used to stretch GPU memory for LLM serving, but its storage behavior has not been characterized at the block-device level. In this paper, we study LMCache through realworld multi-session workloads that span same/different context $\times$ same/different prompt, using over 100 stateless requests per workload. Our comparison is observational rather than factorial: it contrasts two realizable deployment snapshots---LMCache\,0.3.0 with buffered I/O and the kernel page cache, and LMCache\,0.3.16 with prefix-aware deduplication and \texttt{O\_DIRECT}. Key findings define the paper. First, the legacy stack appears almost read-free at the SSD layer in all four workloads, but a \texttt{flush\_ram} experiment shows that this is conditional on page-cache warmth: once the cache is evicted, the same path issues about 1\,GB of sequential reads in 683\,ms. Second, deduplication delivers a dramatic write reduction only for the exact-repeat workload: one set collapses from 1.62\,GB to 11.75\,MB of writes, while the other three sets remain in the 2--5\,GB range. Third, \texttt{O\_DIRECT} converts a hidden and bimodal read cost into an explicit and stable one: on one set, warm TTFT rises from 27\,ms to 93\,ms because each of the 99 warm requests re-reads the same $\sim$12\,MB partial chunk from SSD. The main conclusion is therefore not ``newer is better,'' but a system trade-off: \emph{predictability versus average throughput}.
\end{abstract}


\section{Introduction}
\label{sec:intro}

Modern LLM serving stacks increasingly behave like memory hierarchies. GPU KV cache capacity is the first hard limit, and when contexts or batch sizes grow, systems must spill state to CPU memory, local SSD, or the network. Recent work has improved this stack from many angles: dynamic GPU KV allocation and serving kernels~\cite{vllm,vattention,flashattention,flashattention2}, iteration-level scheduling and batching~\cite{orca,sarathi}, heterogeneous or offloaded inference~\cite{deepspeed,flexgen,neo,specoffload,lamina,inf2}, prefill/decode disaggregation~\cite{splitwise,distserve,adrenaline,memserve}, and structured or stateful runtimes with prompt reuse~\cite{sglang,chunkattention,cacheblend}. Parallel lines of work reduce KV memory itself through eviction, compression, quantization, sparsity, or selective loading~\cite{h2o,streamingllm,snapkv,pyramidkv,kivi,cachegen,infinigen,minference,alisa}. Only recently have disk-mediated KV sharing and compute-vs-load orchestration been studied directly~\cite{sharedragdcache,cake}. Surveys consistently identify efficient inference and KV-cache management as first-order system bottlenecks~\cite{survey_efficient,survey_roofline}.

LMCache sits directly on this fault line. It is now a widely used open-source KV-cache layer for sharing, offloading, and moving KV state across engines and tiers~\cite{lmcache}. In practice, however, operators do not choose LMCache as an abstract API. They choose a \emph{storage path}: buffered I/O plus the kernel page cache, or direct I/O that makes storage traffic explicit. The performance consequences of that choice are still under-characterized.

This paper studies exactly that question using LMCache traces collected from two versions that reflect two realizable deployment points in current systems. The older point, LMCache\,0.3.0, uses buffered file I/O and naturally benefits from the Linux page cache. The newer point, LMCache\,0.3.16, adds prefix-aware deduplication and supports \texttt{O\_DIRECT}. Our main claim is not that one version dominates the other. Instead, the two points implement two different cost models:
\begin{itemize}
\item \textbf{Buffered I/O + page cache} can make SSD reads disappear entirely in the common warm-cache case, but that benefit depends on host memory state outside LMCache's control.
\item \textbf{\texttt{O\_DIRECT} + dedup} makes read cost explicit and stable, but pays that cost on every retrieval, even when the same data would likely have remained hot in RAM.
\end{itemize}

We make three major contributions. First, we provide a block-layer characterization across four stateless workloads that cover exact repetition, fixed-context/varying-prompt traffic, fixed-prompt/varying-context traffic, and fully varied traffic. Second, we use a page-cache-eviction experiment to show that the legacy stack's ``zero reads'' are conditional rather than mechanism-free. Third, we show that the new stack's strongest deduplication win is narrow: it appears only in the exact-repeat workload, while the more production-like varied workloads still write GB-scale data. The result is a clear systems narrative: \emph{LMCache's current SSD offload path trades average throughput for predictability}.

\section{Relation to Prior Work}
\label{sec:related}

Our study sits at the intersection of three research threads. The first is \emph{LLM serving systems}, which optimize batching, memory management, or phase placement to improve latency/throughput trade-offs~\cite{vllm,orca,deepspeed,flexgen,splitwise,sarathi,distserve,sglang,loongserve,lserve,vattention}. These systems establish that inference is often memory-bound, but they usually treat offloaded KV access as an abstract tier rather than characterizing how buffered and direct storage paths differ at the SSD layer.

The second thread is \emph{KV-cache reduction and reuse}. Prior work prunes, compresses, quantizes, or selectively keeps KV state~\cite{h2o,streamingllm,snapkv,pyramidkv,kivi,cachegen,infinigen,minference,alisa}, while systems such as ChunkAttention, CacheBlend, MemServe, and recent disk-based RAG caches aim to improve prompt reuse or cross-request sharing~\cite{chunkattention,cacheblend,memserve,sharedragdcache}. These works show that KV reuse can save substantial compute, but they rarely ask what the storage device actually sees when reuse happens.

The third thread is \emph{compute-versus-load coordination}. Cake explicitly argues that prefill computation and KV loading should be co-optimized rather than treated as separate choices~\cite{cake}, and heterogeneous/offloaded systems similarly balance GPU computation against movement through CPU, SSD, or near-storage tiers~\cite{neo,specoffload,inf2,lamina,adrenaline}. Our results provide a concrete block-level explanation for why such coordination matters: in short-context, partial-chunk regimes, loading the reused KV object can cost more than recomputing the tail it is meant to save.

Relative to these threads, 
our paper investigates a previously unexplored problem and provides valuable insights through comprehensive experimental evaluation.
We do not propose a new serving engine or a new KV policy. Instead, we characterize the \emph{observed storage behavior} of a practical open-source KV-cache layer and show that the dominant trade-off is not simply write reduction, but the shift from hidden read cost to explicit read cost.

\section{Experimental Context and Evaluation Methodology}
\label{sec:method}

\subsection{Two deployment snapshots}

Our comparison is framed as \emph{observational}. The legacy traces were archived from September~2025 using vLLM\,0.9.0 + LMCache\,0.3.0 on a Samsung~980 SSD. The new traces were collected in March--April~2026 using vLLM\,0.11.0 + LMCache\,0.3.16.dev11 on an RTX~5090 (32\,GB) workstation with a Crucial~P310 NVMe SSD. The new buffered-I/O path with \texttt{save\_unfull\_chunk=True} was unstable in our environment and appeared to trigger a use-after-free crash, so we compare two realizable points instead of a full factorial matrix over version, SSD, and I/O mode. We therefore interpret the paper as a \emph{trade-off characterization}, not a causal ablation.

\subsection{Tracing and workloads}

We traced both stacks with \texttt{bpftrace} on \texttt{tracepoint:block:block\_rq\_issue}. This captures I/O \emph{submission} to the block layer rather than completion. Consequently, short microbursts can exceed device bandwidth when measured as submit rate---for example, the new stack submits 96 128\,KB writes in $\sim$3\,ms on Set~A (an apparent $\sim$10.4\,GB/s, well above the PCIe Gen4\,x4 ceiling of $\sim$8\,GB/s)---and we avoid interpreting those spikes as physical throughput.

All requests are \emph{multi-session} and stateless: every request sends a single \texttt{messages=[\{role:user, content:context+prompt\}]} body with no accumulated history. Each scenario runs for 100 requests. The workload matrix is shown in Table~\ref{tab:workloads}. Set~A models regenerate/idempotent-retry behavior; Set~B matches common stateless APIs with a fixed shared context and varying user query; Set~C represents a fixed question applied to changing short snippets; Set~D is an unconstrained baseline. The token ranges in Table~\ref{tab:workloads} are the realized chat-template lengths from our traces. Importantly, all four workloads remain in short-context, partial-chunk territory ($<2$ full 256-token LMCache chunks), which means they stress the treatment of incomplete tails rather than the long-prefix regimes targeted by many prefix-sharing systems.

\begin{table}[t]
\caption{Workload matrix. Each scenario sends 100 independent requests.}
\label{tab:workloads}
\centering
\scriptsize
\begin{tabular}{@{}lccp{0.47\columnwidth}@{}}
\toprule
Set & Pattern & Tokens & Representative use case \\
\midrule
A & same/same & $\sim$94 & regenerate, idempotent retry, repeated benchmark prompts \\
B & same/diff & $\sim$318 & fixed system/context prompt with changing user queries \\
C & diff/same & 150--220 & one classifier/question over many short snippets \\
D & diff/diff & 350--450 & mixed traffic baseline \\
\bottomrule
\end{tabular}
\end{table}

\section{From Zero Reads to Hidden Reads}
\label{sec:hidden}

\subsection{Legacy LMCache looks almost read-free}

Table~\ref{tab:summary} summarizes the measurements. At first glance the legacy stack seems remarkable: across all four workloads, the SSD sees only 0--0.05\,MB of reads at the block layer, even though LMCache logically performs \texttt{get} operations after the first identical request in Set~A. At the same time, writes remain substantial and workload-sensitive, ranging from 1.62\,GB (Set~A) to 3.79\,GB (Set~B). This is consistent with the old path's naive write-through behavior: repeated content is still written back out as chunk files because there is no prefix-aware dedup check.

\begin{figure*}[t]
\centering
\subfloat[LMCache\,0.3.0 + buffered I/O + page cache. Writes are visible; LMCache data reads are almost entirely hidden.]{
\includegraphics[width=0.79\textwidth]{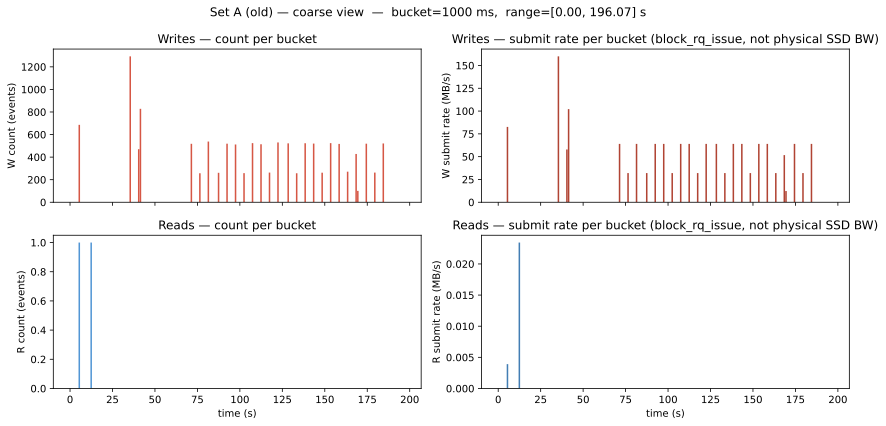}
\label{fig:setA-old}}
\hfill
\subfloat[LMCache\,0.3.16 + \texttt{O\_DIRECT}. Writes collapse to an initial burst, while the 99 warm requests generate recurring reads.]{
\includegraphics[width=0.79\textwidth]{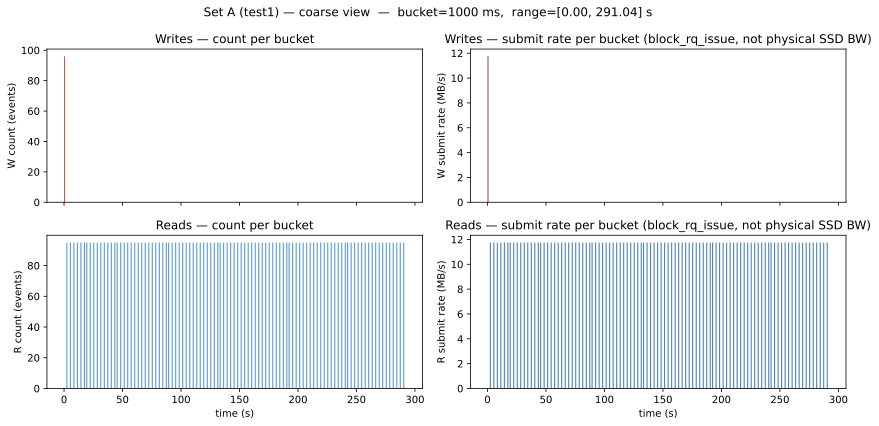}
\label{fig:setA-new}}
\caption{Set~A at 1\,s granularity. The two versions exhibit near-mirror-image I/O patterns.}
\label{fig:setA-mirror}
\end{figure*}

The temporal structure is equally important. Figure~\ref{fig:setA-mirror}(a) shows Set~A under LMCache\,0.3.0. Instead of 100 regular write bursts aligned with requests, writes are spread across the full experiment window and released in irregular spikes by kernel writeback. In other words, buffered I/O obscures application rhythm: the block layer shows when dirty pages happen to be flushed, not when LMCache logically decided to persist a chunk. The few residual reads in the old traces are metadata reads (e.g., ext4 inode/journal activity), not LMCache data fetches. Across Set~A--D, these metadata reads remain tiny compared with writes: the archived traces contain only 14--44 such events per scenario, while write counts range from 13{,}199 to 30{,}994. Notably, the ratio of metadata reads to data writes stays close to $1{:}900$--$1{:}1000$ across all four workloads, which suggests these events track ext4 inode/journaling activity proportional to write volume rather than anything workload-specific. The more meaningful result is therefore qualitative: under a warm page cache, LMCache retrievals are almost invisible to the SSD.

\begin{table}[t]
\caption{Summary across the four workloads. Legacy and new columns report block-layer data I/O totals. Warm TTFT is the mean over requests 2--100 in the new traces; the legacy warm TTFT available from archived logs is 27\,ms for Set~A.}
\label{tab:summary}
\centering
\scriptsize
\begin{tabular}{@{}lccccc@{}}
\toprule
Set & \multicolumn{1}{c}{$W_{\text{old}}$} & \multicolumn{1}{c}{$R_{\text{old}}$} & \multicolumn{1}{c}{$W_{\text{new}}$} & \multicolumn{1}{c}{$R_{\text{new}}$} & \multicolumn{1}{c}{Warm TTFT$_{\text{new}}$} \\
& (GB) & (MB) & (MB) & (MB) & (ms) \\
\midrule
A & 1.62 & 0.03 & 11.75 & 1163.25 & 93.1 \\
B & 3.79 & 0.00 & 3973.88 & 0.00 & 49.5 \\
C & 2.73 & 0.00 & 2030.50 & 0.12 & 40.1 \\
D & 2.61 & 0.05 & 4829.50 & 0.00 & 59.2 \\
\bottomrule
\end{tabular}
\end{table}

\subsection{An eviction experiment reveals the hidden cost}

The natural hypothesis is that the old stack is not truly read-free; rather, its reads are satisfied from the page cache and therefore do not reach the SSD. We validated this directly. We first populated roughly 1\,GB of LMCache data, then forced memory pressure and cache eviction (\texttt{drop\_caches} plus explicit RAM pressure), and finally re-read the same file while tracing the block layer.

\begin{figure*}[t]
\centering
\includegraphics[width=0.9\textwidth]{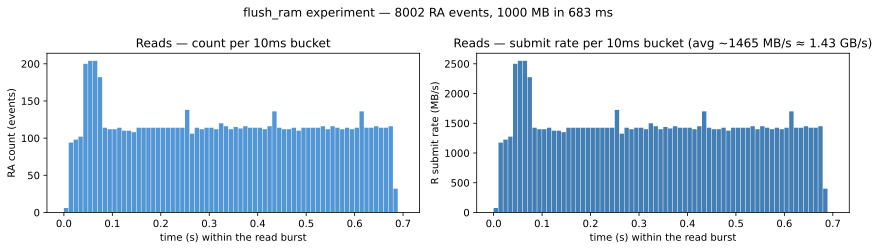}
\caption{\centering The \texttt{flush\_ram} validation experiment. After page-cache eviction, re-reading the same data path generates 8{,}002 128\,KB reads and transfers $\approx$1\,GB in 683\,ms.}
\label{fig:flushram}
\end{figure*}

Figure~\ref{fig:flushram} shows what happened next. The trace contains 8{,}002 128\,KB read-ahead events, totaling $\approx$1\,GB in 683\,ms, for an observed sequential rate of about 1.43\,GB/s. Once the page cache is cold, the ``missing'' reads become a dense sequential scan at the storage layer. This confirms the key point of the section: \textbf{legacy LMCache's zero-read appearance is conditional on RAM residency}. The cost model is therefore bimodal: effectively zero in the warm-cache regime, but suddenly storage-bound once the working set no longer fits or host memory pressure changes.

\section{Current LMCache: Explicit Reads and Narrow Dedup Gains}
\label{sec:new}

\subsection{Set~A becomes predictable and slower}

Set~A provides the clearest contrast between the two stacks. With LMCache\,0.3.16 + \texttt{O\_DIRECT}, writes drop from 1.62\,GB to 11.75\,MB, i.e., about a 138$\times$ reduction. This is the upside of prefix-aware deduplication: only the first request persists the chunk; subsequent identical requests reuse it. But Figure~\ref{fig:setA-mirror}(b) shows the hidden cost of that win. The 99 warm requests trigger 9{,}405 reads and 1.16\,GB of SSD traffic in aggregate, which is exactly 95 reads and 11.75\,MB per warm request.

This explicit retrieve cost also appears in user-visible latency. The archived Set~A logs for the legacy stack report 27\,ms mean warm TTFT, while the new stack reports 93.1\,ms. The $\sim$66\,ms gap is not a small perturbation; it is the dominant warm-path cost of forcing the retrieve through SSD on every request.

\begin{figure}[t]
\centering
\includegraphics[width=\columnwidth]{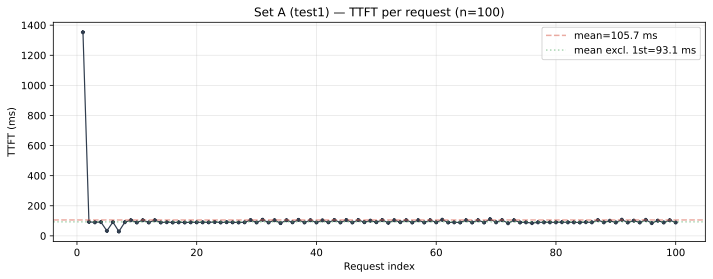}
\caption{Set~A TTFT under LMCache\,0.3.16 + \texttt{O\_DIRECT}. After the first cold request, warm TTFT stays mostly within the 89--105\,ms band, illustrating low variance and a higher latency floor.}
\label{fig:setA-ttft}
\end{figure}

Figure~\ref{fig:setA-ttft} shows that the resulting latency is nevertheless very stable after the first request. In our trace, the 99 warm requests fall mostly into the 89--105\,ms range, with only a few lower outliers early in the run. This is the practical meaning of ``predictability'' in the title: the warm path is slower, but its variance is narrow because it no longer depends on whether the kernel still happens to cache the chunk.

\subsection{Why the read cost is high: partial-chunk amplification}

The most interesting mechanism is a granularity mismatch between two caches. In our setup, vLLM's GPU prefix cache works at 16-token blocks, while LMCache persists 256-token chunk files. Set~A contains only about 94 chat-template tokens, so the first 80 tokens fit into five full GPU prefix blocks, but the remaining 14-token tail does not. LMCache still stores the entire partial chunk as one file. The result is that a warm Set~A request must read the full $\sim$12\,MB chunk file from SSD even though only the tail remains logically useful after GPU prefix matching. Relative to the unmatched 14-token tail, this is about a 6.7$\times$ read amplification.

This also explains an otherwise surprising observation in Table~\ref{tab:summary}: Set~A warm TTFT (93.1\,ms) is \emph{higher} than Set~B warm TTFT (49.5\,ms), even though Set~B recomputes a much longer prompt. In our short-context regime, loading the partial chunk from SSD can cost more than simply recomputing a few hundred tokens on the GPU. Put differently, exact KV reuse is not automatically a latency win when the reused object is too coarse.

\subsection{Set~B/C/D show dedup gains are narrow}

The remaining three workloads tell the second half of the story. Set~B, Set~C, and Set~D write 3.97\,GB, 2.03\,GB, and 4.83\,GB, respectively, under the new stack. These are the same order of magnitude as the legacy results. Their read volume is essentially zero because there is little or nothing to retrieve: prompt or context variation changes the chunk content and therefore the dedup key. In the new logs, these runs correspond to \texttt{LMCache hit tokens: 0} on almost every request.

This matters because Set~B/C/D are not pathological corner cases; they are closer to common stateless production traffic than exact repetition is. Set~B is particularly revealing. Even though all 100 requests share the same short context, the prompt variation is enough to move the full request into a new chunk hash, so the shared prefix is still rewritten. The new trace writes 3.97\,GB in total, or about 39.7\,MB per request, while reading nothing back. In effect, LMCache behaves like a write sink rather than a reuse layer.

Set~C and Set~D complete the picture. Set~C writes 2.03\,GB total ($\sim$20.3\,MB per request) and sees only one 128\,KB read across the whole run, which is too small to represent meaningful KV reuse. Set~D is the worst case: 4.83\,GB written, no reads, and warm TTFT of 59.2\,ms. Together, these runs show that when prompt or context changes at short lengths, deduplication at chunk-file granularity is usually defeated before it can help.

The broader implication is that LMCache's advertised write collapse is not a property of ``having dedup'' in the abstract; it is a property of workloads whose repeated content survives chunking boundaries unchanged. In all three varied workloads, that condition fails. The new stack's most dramatic benefit is therefore a \emph{best-case} benefit, not a broad one.

\section{Discussion}
\label{sec:discussion}

\subsection{The real choice is predictability versus average throughput}

The old and new stacks each solve a real problem, but not the same one. The legacy stack optimizes the common warm-cache case. If host RAM is abundant and the working set remains resident, SSD reads vanish and Set~A warm TTFT stays near 27\,ms. The cost is that this behavior depends on external memory conditions that LMCache does not control. When those conditions change, the \texttt{flush\_ram} experiment shows that the read path can suddenly surface as a GB-scale storage burst.

The new stack removes that external dependency. Every Set~A warm request behaves similarly: about 95 submitted reads, about 11.75\,MB transferred, and about 93\,ms TTFT. This is worse on average in our single-user short-context setting, but more predictable for operators who care about tail latency or service-level objectives. The key insight is therefore not a version ranking. It is a cost-model distinction:
\emph{buffered I/O hides cost opportunistically; direct I/O exposes cost deterministically.}

Table~\ref{tab:costmodel} makes this concrete on Set~A, where the two stacks drive the SSD into opposite regimes from the same user-visible workload.

\begin{table}[t]
\caption{Set~A: two cost models contrasted. The same 100 identical requests produce nearly opposite block-layer behavior.}
\label{tab:costmodel}
\centering
\scriptsize
\begin{tabular}{@{}lll@{}}
\toprule
Dimension & Legacy (page cache) & New (\texttt{O\_DIRECT}) \\
\midrule
Writes & 1.62\,GB (100$\times$) & 12\,MB (1$\times$) \\
Reads & $\sim$0 warm / 1\,GB cold & 1.16\,GB (99$\times$) \\
Read model & bimodal & constant \\
Warm TTFT & 27\,ms (hot) & 93\,ms \\
TTFT variance & RAM-dependent & narrow (89--105\,ms) \\
Schedule visible? & via writeback only & per-request \\
Dedup benefit & none (naive W/T) & yes (Set~A only) \\
Pareto position & unstable & stable but slower \\
\bottomrule
\end{tabular}
\end{table}

\subsection{Implications for KV-cache design}

Our results also interact with recent KV-cache research. Work on context reuse and fusion~\cite{cacheblend,chunkattention,memserve,sharedragdcache,cake} implicitly assumes that loading a cached object is worthwhile once it exists. Our Set~A result refines that assumption: whether loading wins depends on \emph{object granularity} and the boundary between GPU-resident and storage-resident reuse. If the remaining unmatched suffix is tiny, retrieving a whole partial chunk may be worse than recomputation. This suggests two concrete design directions:
\begin{itemize}
\item \textbf{Finer-grained persistence for partial tails.} If LMCache could materialize only the unmatched suffix, it could preserve dedup benefits without paying full-chunk read cost.
\item \textbf{Chunk-aware dedup and layout.} Persisting shared prefixes separately from varying tails would let short stateless API traffic benefit from dedup even when only part of the request is repeated. The expected gain scales with how much of a request is shared: for Set~B's 48-token shared context ($\sim$19\% of a chunk), an aligned layout would save only $\sim$15\% of the writes; but for a chunk-aligned 1024-token shared context typical of RAG, the saving jumps to $\sim$80\%. The mechanism is therefore most valuable exactly where today's LMCache helps least.
\item \textbf{Hybrid compute-or-load policies.} Systems like Cake~\cite{cake} already argue that loading and recomputing should be co-optimized. Our measurements provide a block-level explanation for why that policy is necessary even on local SSDs.
\end{itemize}

\subsection{Operational takeaways}

The measurements suggest a simple decision rule based on two workload properties: \emph{repetition structure} and \emph{memory slack}. If the deployment mostly serves short repeated prompts on a machine with abundant host RAM and little interference, buffered I/O remains attractive because the page cache can eliminate warm reads almost completely. Our legacy Set~A trace is exactly that regime: the SSD sees essentially no data reads and warm TTFT stays low. In contrast, if the operator needs a stable storage budget and stable warm latency independent of host memory pressure, \texttt{O\_DIRECT} is the safer choice because it exposes the true retrieve cost upfront.

A second practical lesson is that ``cache hit'' is too coarse a metric for SSD-backed KV reuse. Set~A does hit, but the hit is expensive because the reusable object is a whole partial chunk. Set~B/C/D mostly miss, but their warm TTFT is still lower than Set~A because there is no retrieve path to pay. For operators, this means that hit rate must be interpreted jointly with \emph{retrieved object size}. A system that reports many hit tokens can still underperform if those hits map to coarse storage objects whose load time exceeds the saved compute.

Finally, the four workloads in Table~\ref{tab:workloads} can be read as a small decision map. Exact repetition (Set~A) flips the I/O pattern from ``always write'' to ``write once, read many.'' Varied traffic (Set~B/C/D) remains write-dominated under both stacks, so the dominant question is whether writes should be buffered and hidden by the kernel or submitted directly and predictably. We expect the crossover point to move for longer contexts, larger shared prefixes, and chunk-boundary-aligned RAG prompts. But even without those extensions, the present results already show that SSD-backed KV reuse should be treated as a \emph{workload-specific} optimization, not as a universally positive switch.



\section{Conclusion}
\label{sec:conclusion}

This paper characterized LMCache SSD offload as a systems trade-off rather than an upgrade story. The legacy buffered-I/O path appears almost read-free only because the kernel page cache hides retrievals; once that cache is cold, the same path issues GB-scale reads. The newer \texttt{O\_DIRECT} path makes storage traffic explicit and stable, but on short repeated prompts it pays that cost on every request, raising warm TTFT from 27\,ms to 93\,ms. Prefix-aware deduplication still matters, but its largest write-saving benefit is confined to exact-repeat traffic and largely disappears once prompt or context varies. The practical takeaway is: for current LMCache SSD offload, the central design question is \emph{whether the deployment values predictable latency more than average warm-path speed}. Resolving the buffered-path instability in the new version is the highest-leverage next step. 

\balance

\end{document}